\documentclass[%
aip,
amsmath,amssymb,
reprint,%
]{revtex4-2}
\usepackage{graphicx}
\usepackage{dcolumn}
\usepackage{bm}

\usepackage[utf8]{inputenc}
\usepackage[T1]{fontenc}
\usepackage{mathptmx}
\usepackage{etoolbox}

\makeatletter
\def\@email#1#2{%
	\endgroup
	\patchcmd{\titleblock@produce}
	{\frontmatter@RRAPformat}
	{\frontmatter@RRAPformat{\produce@RRAP{*#1\href{mailto:#2}{#2}}}\frontmatter@RRAPformat}
	{}{}
}%
\makeatother
\begin{document}
	
	\preprint{AIP/123-QED}
	
	\title[Linear MR in YSi]{Linear unsaturated magnetoresistance in YSi single crystal}
   \author{Vikas Saini}
	\author{Souvik Sasmal}
	\author{Ruta Kulkarni}
	\author{Arumugam Thamizhavel}
	\email{thamizh@tifr.res.in}
	\affiliation
	{Department of Condensed Matter Physics and Materials Science, Tata Institute of Fundamental Research, Homi Bhabha Road, Colaba, Mumbai 400 005, India.}
	
	
	\date{\today}
	
	\begin{abstract}
		
		Linear magnetoresistance is a phenomenon that has been observed in a few topological compounds that originate from classical and quantum phenomena.  Here, we performed electrical transport measurements, in zero and applied magnetic fields, on the YSi single crystal along all three principal crystallographic directions of the orthorhombic crystal structure.   For $I~\parallel~[001]$ and $H~\parallel~[100]$ direction above $\approx 10$~T, mobility fluctuation driven linear magnetoresistance is observed without any sign of saturation up to $14$~T magnetic field. Anisotropy in the Fermi surface is immanent from the angular dependence of the magnetoresistance. Kohler rule violation is observed in this system and Hall data signifies multiple charge carriers in YSi.
		
	\end{abstract}
	
	\maketitle
	
	%
	
	
	In metallic systems, the resistance caused by scattering centers in applied magnetic fields is defined as magnetoresistance (MR) and is generally expressed as [$\frac{R(H,T)-R(0, T)}{R(0, T)}]\times100$.  Because of its wide range of possible uses in modern day electronic devices such as magnetic sensors and storage devices, the MR has been intensively studied~\cite{parkin2003magnetically,ali2014large, soh2002making, fert2008nobel}.   In semiclassical theory, the quadratic dependence of MR in low field regime ($\omega \tau~<<1$) can be explained thoroughly which gets saturated in the high field regime. The linear MR can not be explained using classical Boltzmann theory, which was discussed by Kapitza for the first time in 1928 for open Fermi surface systems~\cite{hu2008classical,kapitza1928study}. It has been found that the narrow gap semiconductors such as Ag$_{2}$(Se,Te), MnAs, GaAs obey the classical description of magnetoresistance and a modest amount of doping in intrinsic semiconductors lead to the unsaturated linear magnetoresistance~\cite{willardson1960magnetoresistance, halbo1968magnetoresistance, xu1997large, johnson2010universal,von2009linear, hu2005current}.  Apart from these disordered semiconductors, linearity in magnetoresistance is observed for semimetallic systems $n$-doped Cd$_{3}$As$_{2}$, PtBi$_{2}$~\cite{narayanan2015linear,yang2016giant}. To explain the linear MR, Parish and Littlewood proposed a classical model in that it has been demonstrated that the inhomogeneity in the crystal can lead to linear MR.  The disorder driven linear MR is associated with the average mobility $\mu$ and the width of the mobility disorder $\Delta \mu$.  For $\frac{\Delta \mu}{\mu} < 1$,  high mobility of charge carriers results in the quadratic to linear crossover at the cross-over field $B_c$  $\sim $ $\mu^{-1}$, where $B_c$ increases with temperature, in the second scenario of $\frac{\Delta \mu}{\mu} > 1$, where the width of the mobility distribution is dominating the average mobility results in $B_c$ $\sim$  ${\Delta \mu}^{-1}$.  Hence, when $B_c$ is proportional to $\frac{1}{\mu}$ or $\frac{1}{\Delta \mu}$ linear magnetoresistance is usually observed~\cite{narayanan2015linear,parish2003non}. 
	
	On the contrary, the quantum linear magnetoresistance has been reported in Bi thin films, multilayer epitaxial graphene, Bi$_{2}$Te$_{3}$ nanosheets, Bi$_{2}$Se$_{3}$ thin film, TmB$_{4}$, CaMnBi$_{2}$, SrMnBi$_{2}$  etc.~\cite{lee2002band, yang1999large, hu2007nonsaturating,mitra2019quadratic, wu2015large, wang2017disorder, friedman2010quantum, abrikosov1998quantum, wang2012room,wang2012two,park2011anisotropic}. 
	The detailed study of  quantum magnetoresistance is proposed by Abrikosov.  In the extreme quantum limit, when the system is in the lowest Landau level and if the carrier density and temperature obey following inequalities $n_{0}~\ll~(\frac{e H}{\hbar c})^{\frac{3}{2}}$ and $T~\ll~(\frac{e H \hbar}{m^{*} c})$,  resistivity tensors are enforced to be linear~\cite{abrikosov2000quantum}, $\rho_{xx}~=~\rho_{yy}~=~ \left(\frac{N_{i}H}{\pi n_{0}^2 e c }\right)$ where $N_{i}$ is scattering density.
	
	In this work, we performed the resistivity and magnetoresistance studies on YSi single crystal along the three principal crystallographic directions of the orthorhombic crystal structure.  We observed that the MR is not showing any signature of saturation up to a magnetic field of $14$~T.  Particularly, $J~\parallel$~[001] and $B~\parallel$~[100] direction reveals the linear behaviour in magnetoresistance above $10$~T magnetic field.  From a detailed analysis, we attribute this linear magnetoresistance to the mobility fluctuations and Abrikosov quantum model does not explain the linearity in the MR.  To support this reasoning, here we recall our recent dHvA studies on YSi where we found that the lowest Landau level has not been achieved until an applied magnetic field of $14$~T.  Furthermore, we also found that Kohler's rule is violated in this compound as the MR curves deviated from each other because temperature dependent inelastic scatterings are more dominant than the elastic scatterings.  Angular dependence of the magnetoresistance for $J~\parallel$~[100] and $J~\parallel$~[010] revealed two fold anisotropy in the Fermi surface.  With increasing temperature, MR is found to decrease as the mobility drops which is well correlated with the Hall measurements.
	
	The crystal growth and preliminary analysis of YSi has been reported in our earlier work~\cite{saini2021fermi}.  Briefly, YSi crystallize in the orthorhombic crystal structure with non-symmorphic  space group $Cmcm~(\#63)$ and the single crystal has been grown by Czochralski method in a tetra-arc furnace.  Well defined Laue diffraction spots ascertained the good quality of the single crystal and samples were cut along the three principal crystallographic directions for anisotropic electrical resistivity studies.   Resistivity measurements were performed  by four probe method with the electrical contacts made with $50~\mu$m diameter gold wire using two component epoxy paste. Electrical transport measurements were carried out in the temperature range $1.8 - 300$~K in Quantum Design Physical Property Measurement System (QD-PPMS) equipped with a  $14$~T magnet.  
	
	\begin{figure}[!]
      	\includegraphics[width=0.45\textwidth]{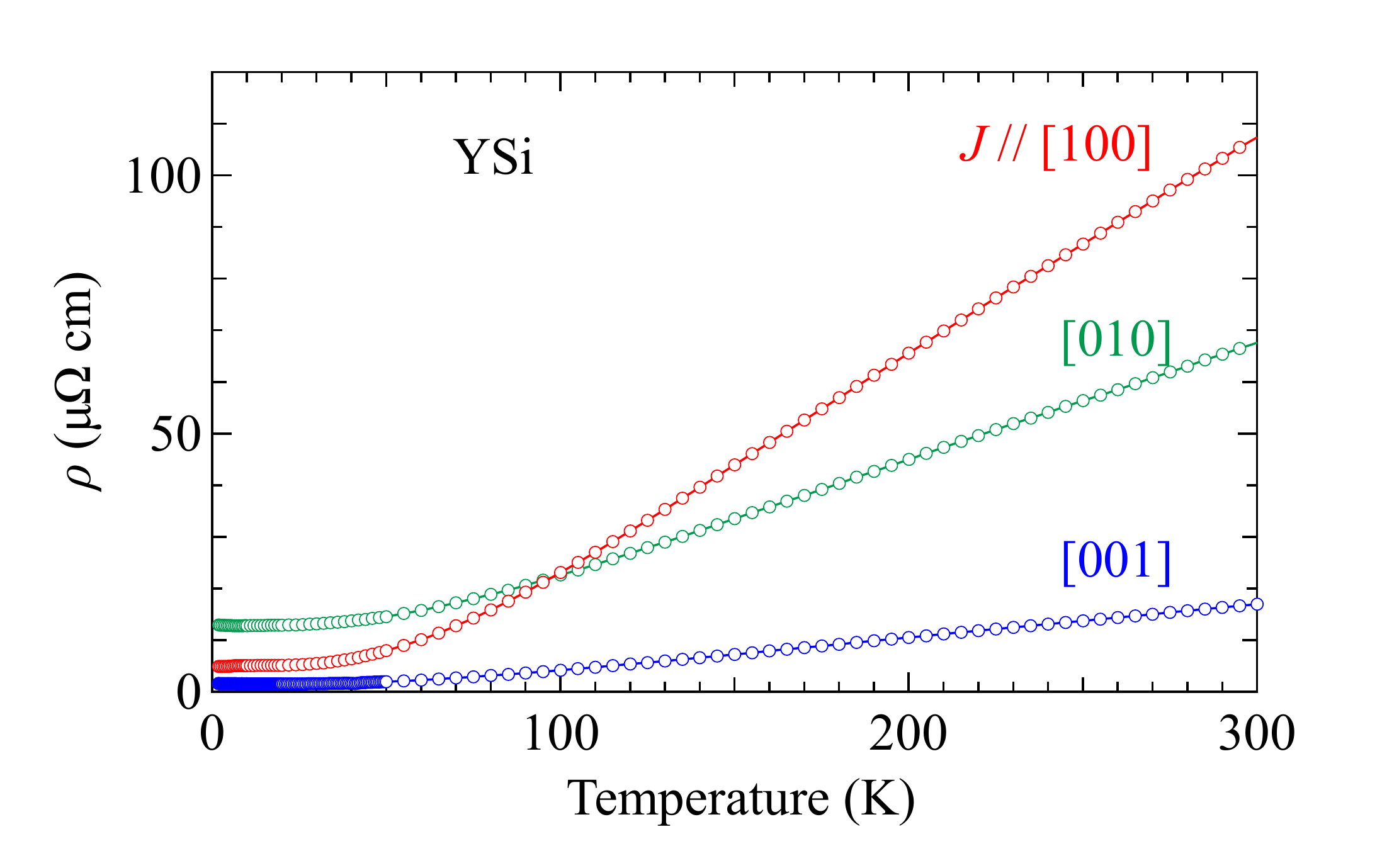}
		\caption{(a) Temperature dependence of electrical resistivity of YSi along the three principal crystallographic directions in the temperature range $1.8 - 300$~K.  }
		\label{Fig1}
	\end{figure}
	
	Figure~\ref{Fig1} depicts the temperature dependence of electrical resistivity of YSi along the three principal crystallographic directions.  The electrical resistivity is typical of a metallic system which decreases as the temperature is decreased.  At room temperature, there is a significant anisotropy in the electrical resistivity with $\rho (300$)~K~$\approx$~107~$\mu \Omega$~cm  for $J~\parallel$~[100] and for the other two directions viz., [010] and [001]-directions it was found to be about  $68$ and $17~\mu \Omega$cm, respectively.  The RRR value estimated from $\rho(300$~K)/$\rho(2$~K) amounts to 22, 5 and 11, respectively for  $J~\parallel$~[100], [010] and [001]-directions.  When magnetic field is applied in the transverse direction, the temperature dependence of electrical resistivity did not show any turn-on and plateau behaviour at low temperature, in all the three directions, as observed in other semimetallic compounds like WTe$_2$,  MoSi$_2$, WSi$_2$ etc~\cite{wang2015origin, matin2018extremely, mondal2020extremely}.
	
	\begin{figure*}[!]
		\includegraphics[width=0.95\textwidth]{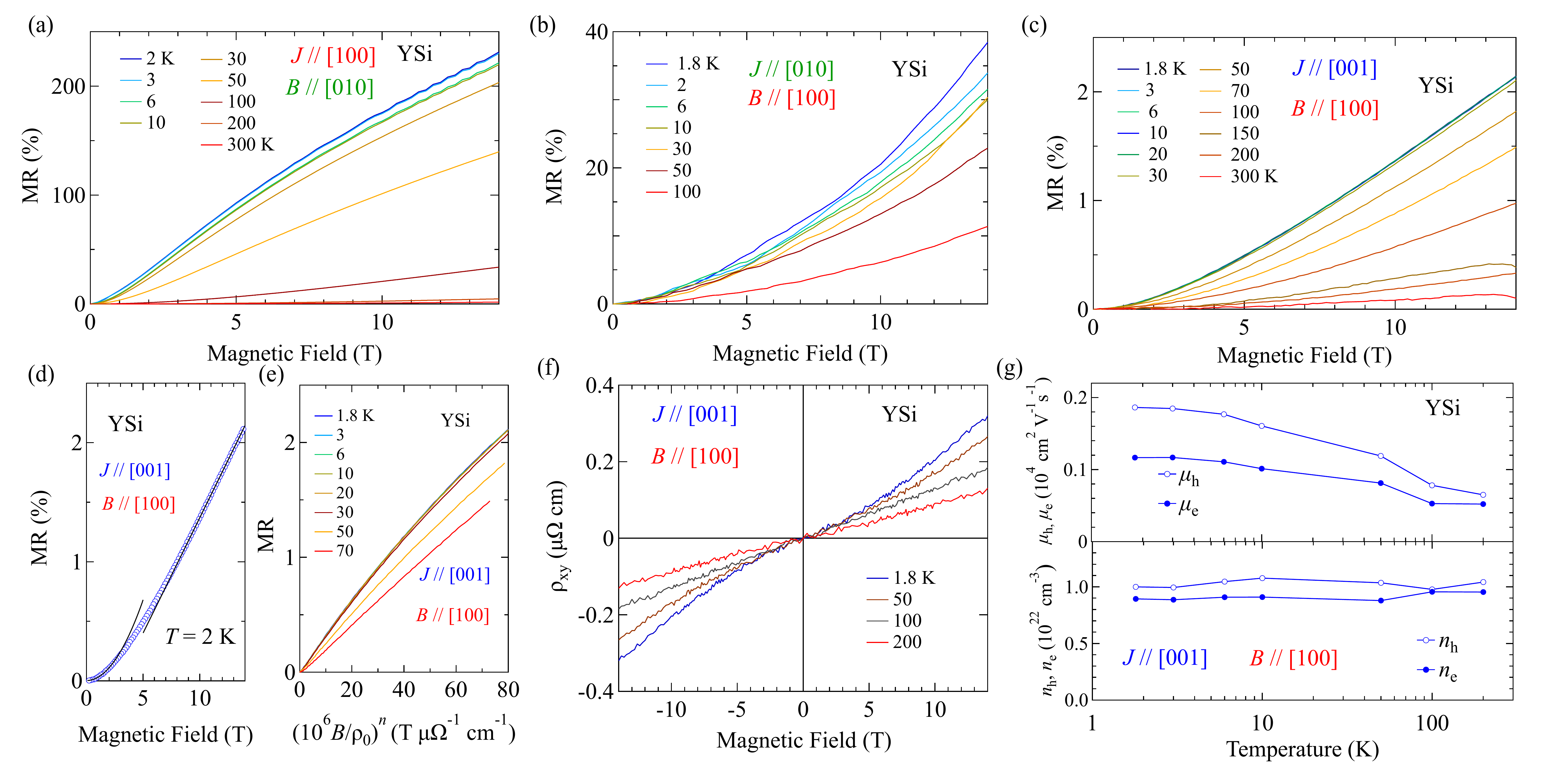}
		\caption{(a-c) MR decreases with increasing temperature for $J~\parallel~[100]$, $J~\parallel~[010]$, and $J~\parallel~[001]$. (d) For $J~\parallel~[001]$ and $B~\parallel~[100]$ quadratic dependence of MR is fitted in the low field regime and in the high field regime linear MR is fitted. (e)  Kohler plot for  $J~\parallel~[001]$ and $B~\parallel~[100]$ (f)  For $J~\parallel~[001]$ and $B~\parallel~[100]$ Hall resistivity for various temperatures. (g) Electron and hole mobilities are suppressing with temperature dependent scatterings and carrier density plot on the logarithmic scale of temperature.}
		\label{Fig2}
	\end{figure*}
	
	For current along the three principal crystallographic directions and magnetic field perpendicular to it, the normalised MR computed from its general definition, for various isothermal temperatures, is illustrated in Fig.~\ref{Fig2}(a), (b) and (c).  At the lowest temperature $1.8$~K, The MR is largest for $J~\parallel$~[100], amounting to $230\%$ in a magnetic field of $14$~T while it is considerably smaller, amounting to 38$\%$ and 2.14$\%$ respectively, for current along the other two directions depicting a large anisotropy.  The MR does not show any sign of saturation up to a magnetic field of $14$~T.  The signature of Shubnikov de Haas oscillations are observed at the lowest temperature in high magnetic fields, indicating a good quality of the single crystal.  The MR decreases as the temperature is increased.  The temperature rise intensifies the scattering rate, as a consequence, the relaxation time ($\tau$) drops down; therefore, the mobility suppresses with increasing temperature, which eventually decreases MR.  At low temperatures, up to $20$~K the decrease in MR is subtle while for temperature greater than $20$~K there is a rapid decrease in the MR which is correlated with the mobility plot as discussed later.
	
	Figure~\ref{Fig2}(d) depicts the field dependence of the transverse MR for $J~\parallel$~[001] and $B~\parallel$~[100] direction at $T = 1.8$~K. The MR can be explained by a linear combination of quadratic and linear field dependence.  The solid lines in Fig~\ref{Fig2}(d) shows the fits to quadratic field dependence at low fields and a linear field dependence at high magnetic fields.  The low field $(0 - 2$~T) quadratic behaviour can be understood by the classical Boltzmann's theory originating from the Lorentz force.  As the applied magnetic field increases the curvature changes and beyond $10$~T the linear behaviour of MR is observed.  Next, we used the Kohler scaling law to analyse the MR data of $J~\parallel$[001] and $B~\parallel$ [100] in the field range $0-5$~T.   Kohler's rule states that the MR  can be described by a scaling function of the applied magnetic field $B$ and mean scattering time  $\tau$ of the conduction electrons.   $\tau$ is inversely proportional to the residual resistivity $\rho_0$ and hence Kohler's scaling law is given by the expression ${\rm MR} = \alpha (B/\rho_0)^n$, where $\alpha$ and $n$ are constants~\cite{wang2015origin} and should fall on a single curve.  Here, $n$ is a sample dependent parameter that indicates the degree of compensation.  For a perfectly compensated semimetal $n = 2$.  Figure~\ref{Fig2}(e) depicts the Kohler plot where the MR data has been scaled as per  Kohler's scaling expression defined above.  The value of $n$ obtained from scaling the data is 1.6 and it is evident that MR data falls onto a single curve for temperature up to $20$~K and beyond that the Kohler's rule is violated.  This gives evidence for the anisotropic scattering of $\tau$ and does not have the same $T$-scaling from different regions of the Fermi surface.  Also, this suggests more than one type of charge carrier is present in YSi, as discussed in the Hall data below.
	
	\begin{figure}[!]
		\includegraphics[width=0.5\textwidth]{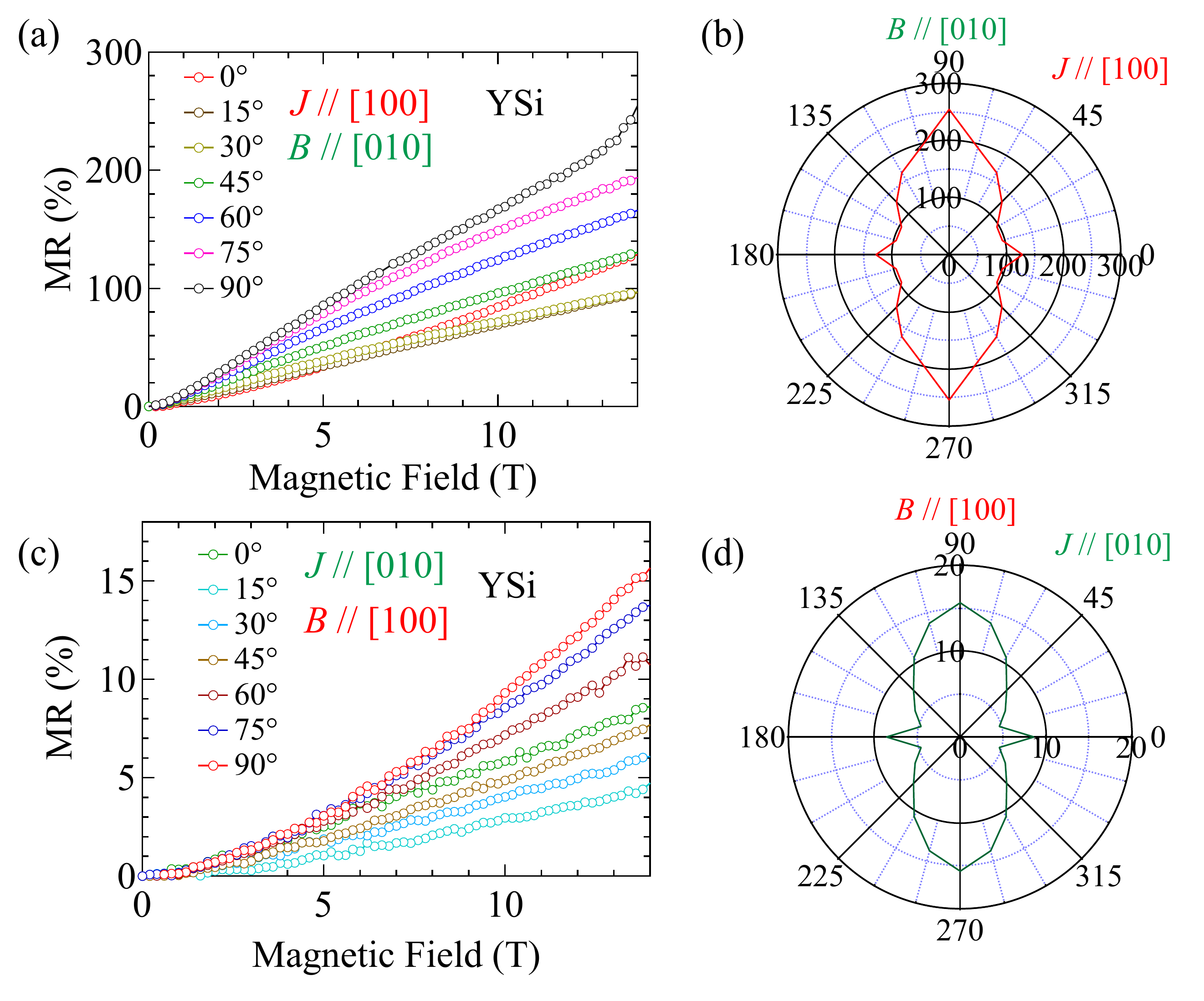}
		\caption{(a) Angular dependence of MR where the magnetic field is rotated from $J~\parallel~[100]$, ($\theta~=~0^\circ$) to $[010]$ direction($\theta$=$90^\circ$). (b) Polar diagram for the (a) plots. (c). Angular dependence of MR for  $J~\parallel~B~\parallel~ [010](\theta~=~0^\circ$) to [100] ($\theta~=~90^\circ$). (d) Anisotropic polar plot for (c) plots which reveals the anisotropic Fermi surface.}
		\label{Fig3}
	\end{figure}
	
	We performed Hall resistivity measurements at different temperatures to further study the electrical transport properties; typical data at a  few specific temperatures are provided in Fig.~\ref{Fig2}(f). The Hall resistivity exhibits positive value for the entire field range up to $14$~T indicating holes are the majority carriers.  At low fields, the Hall resistivity shows a small curvature while at high magnetic fields it is linear.  The non-linearity in Hall resistivity signals the presence of multiple charge carriers in YSi, thus validating our Kohler plot analysis.  Furthermore, this is consistent with our band structure calculations, which revealed that both electron and hole pockets constitute the Fermi surface~\cite{saini2021fermi}.  We now employ the classical two band transport model for analysing the Hall data for $J~\parallel$~[001] and $B~\parallel$~[100].  The Hall resistivity ($\rho_{xy}$), based on the two band model~\cite{gao2017extremely} is defined as : 
	
	\begin{equation}
		\rho_{xy} = \frac{B}{e} \frac{(n_{h} \mu_{h}^{2}-n_{e} \mu_{e}^{2}) + (n_{h}-n_{e})(\mu_{h}\mu_{e})^2 B^2}{(n_{h} \mu_{h}+n_{e} \mu_{e})^2+(n_{h}-n_{e})^2(\mu_{h}\mu_{e})^2 B^2 },
		\label{Eq1}
	\end{equation}
	
	where $n_e$ and $n_h$, are the carrier concentration of electrons and holes and $\mu_e$ and $\mu_h$ are their corresponding mobilities.  The extracted mobility and the carrier concentration are plotted in Fig.~\ref{Fig2}(g).  The hole mobility and hole carrier density are dominating than the electron mobility and carrier density in the entire temperature range.   The numeric values of mobilities at 1.8 K for hole carrier $\mu_{h}$ $=$ 0.18 $\times$ $10^4$~cm$^{2}$V$^{-1}$s$^{-1}$ and for electron carrier is $\mu_{e}$ $=$ 0.11~$\times$~$10^4$ cm$^{2}$V$^{-1}$s$^{-1}$.   The estimated carrier densities are almost temperature independent which correlates well with the pure metallic behaviour of YSi as observed in Fig.~\ref{Fig1}(a) where $\frac{dR}{dT}<0$ is not seen in any temperature interval. 
	
	The angular dependence of MR in increments of $15^{\circ}$ measured at $1.8$~K for the $J~\parallel~[100]$ and [010] directions are shown in Fig.~\ref{Fig3}(a) and (b).  In the polar plot, $0^{\circ}$ corresponds to the current and magnetic field in the parallel orientation, and at $90^{\circ}$ orientation it represents the current and magnetic field in the perpendicular geometry.  For $J~\parallel~[100]$, the MR begins with a slight bend, then steadily grows with increasing angles until $\theta = 90^{\circ}$, where $B$ is parallel to [010]. The highest MR is for $\theta = 90^{\circ}$. Similarly, a modest spike is observed in MR in the other orientation for $J~\parallel~[010]$, following which it progressively increases until $\theta = 90^{\circ}$, as indicated in the polar plot of Fig.~\ref{Fig3}(b). In both the orientations MR shows a two-fold anisotropy, which correlates well with the anisotropic Fermi surface as described in Ref.~\onlinecite{saini2021fermi}.

	\begin{figure}[h]
		\includegraphics[width=0.45\textwidth]{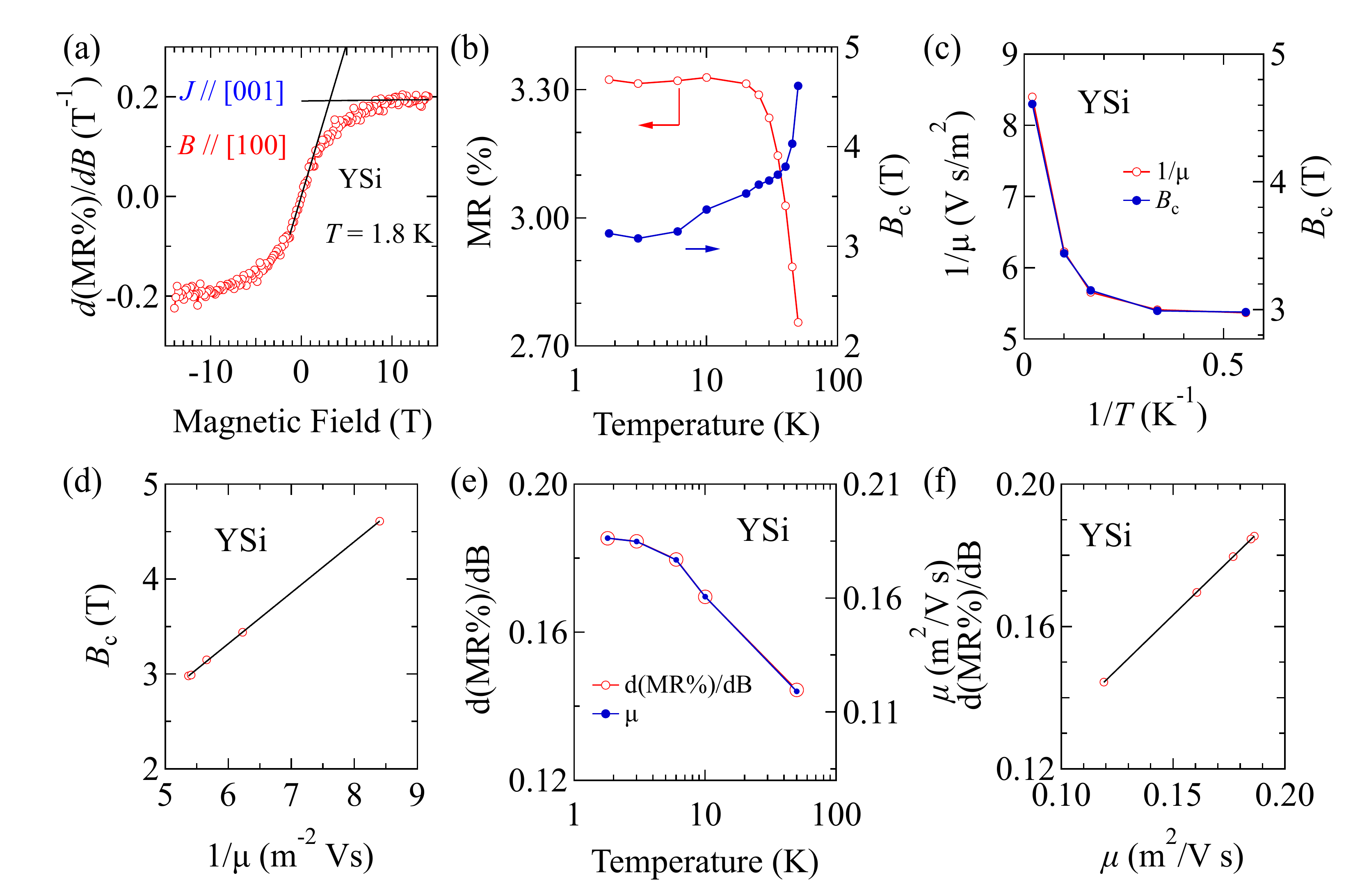}
		\caption{(a) Derivative of MR for  $J~\parallel~c$  and $H~\parallel~a$. (b) Crossover magnetic field and MR temperature dependencies. (c) Crossover magnetic field $B_{c}$ vs inverse mobility $\frac{1}{\mu}$ plot with temperature. (d) Linear behaviour of crossover field with inverse mobility $\frac{1}{\mu}$. (e) Temperature dependencies of the derivative of magnetoresistance and mobility. (f) Linear plot for the $\frac{d(MR\%)}{dB}$ with mobility ($\mu$).}
		\label{Fig4}
	\end{figure}

	Linear magnetoresistance to an extent can be understood by employing classical and quantum models. However, it has been marked that the linear band dispersion in the vicinity of Fermi energy level precipitate to ultrahigh mobility of charge carriers as investigated for Cd$_{3}$As$_{2}$ (9~$\times~10^6$~cm$^{2}$V$^{-1}$s$^{-1}$ at $5$~K) and for NbP ($5~\times~10^6$~cm$^{2}$V$^{-1}s^{-1}$ at $1.85$~K), which do not obey either of the models of linear MR ~\cite{shekhar2015extremely, liang2015ultrahigh}. On the other hand, Abrikosov proposed that in the extreme quantum limit $\hbar\omega~\gg~$ $E_{\rm F}$, where $\omega$ is the cyclotron frequency and $E_{\rm F}$ is the Fermi energy, charge carriers occupy in the lowest degenerated landau level enforce the resistivity tensors to be linear. For YSi, in our previous work, we have shown that even for the smallest pocket, the lowest Landau level is not achieved even upon applying the magnetic field of $14$~T~\cite{saini2021fermi}.
	
		The classical PL model hypothesizes inhomogeneity in the crystal attributes to the linearity in magnetoresistance. If average mobility is dominant than the width of mobility distribution $\frac{\Delta \mu}{\mu}~<~1$, the quadratic to linear crossover field is defined as $B_{c}$ $\sim $ $\frac{1}{\mu}$. Based on the PL model prescription, the crossover field ($B_c$) and the derivative of magnetoresistance (dMR/dB) obey the following relations with mobility $\left( B_c~\propto~\frac{1}{\mu}\right)$ and $\frac{d(MR)}{dB}~\propto~\mu$.  To analyse the linear MR based on PL model first we determined the crossover field $B_c$.  For this purpose, we plotted the $\frac{d(MR\%)}{dB}$ against the applied magnetic field and a representative plot for $T = 1.8$~K is shown in Fig.~\ref{Fig4}(a).  The intersection of the quadratic and linear regime gives the value of $B_c$.  As the temperature increases, the mobility decreases, this causes an increase in $B_c$ at the same time a decrease in MR, this observation is shown in Fig.~\ref{Fig4}(b). From the Hall data, we found that the concentration of hole carriers is dominant in the entire temperature range and hence we considered the mobility of hole carriers for this analysis.  The inverse mobility ($1/\mu$) and $B_c$ when plotted against inverse temperature they exhibited an identical behaviour as shown in Fig.~\ref{Fig4}(c). The PL model has predicted that $B_c$ is proportional to ($1/\mu$), this is well documented in Fig.~\ref{Fig4}(d).   The derivative of MR at a constant magnetic field ($14$~T) decreases with increase in temperature as shown in Fig.~\ref{Fig4}(e), the temperature dependence of mobility is shown in the right axis of Fig.~\ref{Fig4}(e) which follows the same trend as that of the derivative of MR suggesting the linearity in d(MR\%)/dB vs. $\mu$ plot as shown in Fig.~\ref{Fig4}(f).
	
	In summary, we performed electrical transport measurements on a single crystal of YSi. Non-saturating linear MR probed which is originated by mobility fluctuations and the linearity in MR follows the PL model. Hole dominating behaviour has been observed with hole mobility  $\mu_{h}$ $=$ 0.18 $\times$ $10^4$ $cm^{2}V^{-1}s^{-1}$ consistent with the band structure calculations. Kohler's rule is violated in YSi, indicating that inelastic scatterings dominate other than elastic scatterings and multiple carriers.

	\nocite{*}
	\bibliography{references}

	
\end{document}